\documentclass[a4paper,11pt]{article}

\usepackage{eurosym}
\usepackage{fontenc}
\usepackage[utf8]{inputenc}
\usepackage[english]{babel}
\usepackage{amsfonts}
\usepackage{amsmath}
\usepackage{amssymb}
\usepackage{amsthm}
\usepackage{authblk}

\newtheorem{theorem}{Theorem}
\newtheorem{definition}{Definition}

\newtheorem{remark}{Remark}
\newtheorem{example}{Example}

\title{A unified algorithm for colouring graphs of bounded clique-width}
 \author[1]{Bruno Courcelle 
 }
 \author[2]{Ir\`ene Durand
 }
 \author[3]{Michael Raskin \thanks{
 The author has received funding from the European Research Council (ERC)
 under the European Union's Horizon 2020 research and
 innovation programme under grant agreement No 787367 (PaVeS). }
 }
 
 \affil[1]{University of Bordeaux%
 \\\texttt{bruno.courcelle@u-bordeaux.fr}%
 \\ORCID: 0000-0002-5545-8970}
 \affil[2]{University of Bordeaux%
 \\\texttt{irene.durand@u-bordeaux.fr}%
 \\\mbox{ORCID: 0000-0002-5171-7234}}
 \affil[3]{Technical University of Munich%
 \\\texttt{raskin@\{in.tum.de,mccme.ru\}}%
 \\\mbox{ORCID: 0000-0002-6660-5673}}

\begin{document}

\maketitle

\begin{abstract}
Clique-width is one of the graph complexity measures leading to polynomial special-case algorithms for generally NP-complete problems, e.g. graph colourability. The best two currently known algorithms for verifying $c$-colourability of graphs represented as clique-width terms are optimised towards two different extreme cases, a constant number of colours and a very large number of colours. We present a way to unify these approaches in a single relatively simple algorithm that achieves the state of the art complexity in both cases. The unified algorithm also provides a speed-up for a large number of colours.
\end{abstract}

%\keywords{ Clique-width \and Graph colouring \and Parameterised complexity }

%\vspace{1ex}
%\textit{This paper is not eligible for student paper awards}

\section{Introduction}

Clique-width is one of the graph complexity measures leading to polynomial
special-case algorithms for generally NP-complete problems. 
Roughly speaking, clique-width considers binary trees with 
the vertices of the original graph
being the leaves,
and measures how many different kinds of vertices 
can there be in some subtree
from the point of view
of edges to the vertices outside the subtree.
Although
computing the clique-width of a graph is NP-hard \cite%
{Fellows2009CliqueWidthIN}, one might have a good enough clique-width
representation of a graph, be it from heuristics (like those implemented in
TRAG \cite{TRAGWeb}), algorithms with weaker guarantees \cite%
{Bodlaender2010FasterAO}, special algorithms for restricted graph classes
(like graphs with few $P_4$ subgraphs \cite{DBLP:journals/ijfcs/MakowskyR99}%
), or even a precise computation (using, for example, SAT solvers \cite%
{DBLP:journals/tocl/HeuleS15}) as a per-graph investment to be reused for
computing multiple properties of the graph.

One of NP-complete graph problems is graph colourability.
While there are multiple notions of graph colouring, 
in the present paper we consider only assignments of colours to vertices
with the ends of each edge having different colours.
 There are multiple
known results on complexity of $c$-colourability for graphs with a known
decomposition witnessing a "low"  clique-width .

Most of them are based on a natural refinement of the algorithm implied by
Courcelle's theorem for colourability with a fixed number of colours; the
same algorithm is sometimes described as a straightforward dynamic
programming approach. Specifically, these algorithms traverse the subterms
of a  clique-width  term 
(a formal representation of a witness of low clique-width)
 from the leaves towards the root, and for each
subterm compute the set of possible colourings of the subgraph corresponding
to the subterm. The vertices sharing a label are tracked together. For a
fixed  clique-width and a fixed number of colours the time per node of the
syntax tree of the term is constant.

From the point of view of number of colours there are two limit cases. On
the one hand, the colourability problem is already NP-complete for $3$
colours. On the other hand, when computing the chromatic number in the
general case, the number of colours could in principle be as large as $n$,
and one could want to limit the impact of the number of colours on
complexity. Both these cases have been studied; a fine-grained tight
(assuming exponential time hypothesis) bound has been obtained in the former
case \cite{DBLP:conf/icalp/Lampis18}, and an algorithm polynomial in $n$ for
each fixed clique-width (with a degree of the polynomial exponential in the
number of colours) has been obtained in the latter \cite{Kobler2003EdgeDS,%
DBLP:conf/wg/MakowskyRAG06}. For the latter case there is also a lower bound
assuming exponential type hypothesis \cite{Golovach2018CliquewidthIT}.
However, the algorithms for these two cases are presented from different
points of view. This follows naturally from the number of labels being much
larger than the number of colours in the one case but much smaller in the
other. Thus it remained unclear whether one needs to choose 
which of the two approaches to use in advance.

In the present paper we describe both sets of optimisations inside a unified
framework which allows their simultaneous application. Reformulating the
optimisations from the literature and applying them at once yields a single
relatively simple algorithm verifying $c$-colourability, that naturally
achieves the state-of-the-art complexity in both extreme cases. The
reformulation is not purely formal as care needs to be taken to make sure
each optimisation does not lead to overhead outside of the original area of
application. We believe that a uniform representation of the two
optimisations side by side makes their corresponding core ideas clearer. The
simultaneous application of the optimisations leads to a moderate
improvement to the state-of-the-art complexity in case of large number of
colours, reducing the constant factor in the degree of polynomial
(which is itself exponential in the number of colours).

%% We also evaluate how close these optimisations are to exhausting
%% the possibilities of reducing the number of colouring configurations 
%% to consider.
%% We present a modified algorithm that preserves
%% the upper bounds on complexity in both limit cases,
%% and minimises the set of options for each subterm 
%% conditional on the context
%% information about the rest of the term
%% of the form
%% used in the literature on clique-width based colouring \cite{DBLP:conf/icalp/Lampis18}
%% and on general improvements to algorithms arising 
%% from Courcelle's theorem \cite{DBLP:journals/japll/CourcelleD12}.

The rest of the paper is organised as follows. In the next section we remind
the definitions related to clique-width decomposition of graphs. In the
following section we present our main result and discuss its relation to the
previously known results. Next we describe the details of the algorithm. The
following two sections contain the proof of correctness and complexity
bounds. We finish with a brief conclusion.

\section{Graphs and clique-width}

Graphs are finite, undirected without loops and multiple edges. A vertex may
have a \emph{label} in $[k]:=\{1,...,k$\} (the graph is $k$-\emph{labelled})
and/or a \emph{colour} in $[c]$ (the graph is $c$-\emph{coloured}).
Of course the choice of specific finite sets of labels and colours only
matters when talking about specific graphs and specific colourings,
i.e. in the examples.

Certain $c$-coloured labeled graphs can be constructed from basic graphs~$%
\boldsymbol{a}$ where $a\in{}[k]$, and the following operations:

\begin{itemize}
\item $\oplus$ constructing the union of two disjoint graphs,

\item the unary operation $add_{a,b}$ for $a,b\in \lbrack k]$, $a<b$ that
adds an edge between each $a$-labelled vertex and each $b$-labelled vertex
(unless they are adjacent),

\item the unary operation $relab_{a\rightarrow b}$ changes every vertex
label $a$ into $b$.
\end{itemize}

The \emph{type }of $G$, denoted by $\pi (G),$ is the set of labels of its
vertices.

From these operation symbols, we can build terms that denote $k$-labelled
graphs. 
When discussing a term and the graph it represents,
we often want to refer to specific vertices regardless of their labels,
which can coincide with each other
 or vary depending on the subterm in question.
In this case we use the notation $\boldsymbol{1}(x)$ where $x$ is 
an arbitrary vertex name.
We expect vertex names to be unique. An example is:

\begin{quote}
$t=add_{1,2}(relab_{3\rightarrow 1}(add_{1,3}(\boldsymbol{1}(w)\oplus 
\boldsymbol{3}(w^{\prime }))\oplus \boldsymbol{2}(x)\oplus add_{1,3}(%
\boldsymbol{1}(y)\oplus \boldsymbol{3}(z))))$
\end{quote}

that denotes the graph $G$ obttained from the path $w-w^{\prime }-x-y-z$
by addition of the edge $w-x$, where the labels of $w,w^{\prime },x,y,z$ are
respectively $1,1,2,1,3$ in the nullary symbols that create them. We call
such terms \emph{clique-width terms}, but as we do not consider any other
kinds of terms, we omit the \textquotedblleft
clique-width\textquotedblright{} in the present paper. Each term denotes a
vertex labelled graph $\boldsymbol{val}(t)$ whose vertices are those
specified by the nullary symbols of $t$. (No two occurrences of nullary
symbols denote the same vertex.)  The term's \emph{width} is the number of
labels that occur in $t$.

%% A term over the above defined operations is \emph{well-formed} if no two
%% occurrences of nullary symbols denote the same vertex (so that the graphs
%% defined by the two arguments of any operation $\oplus $ are disjoint). We
%% call them the \emph{clique-width terms}, or simply \emph{terms} in the
%% sequel. Each term $t$ denotes a vertex labelled graph $\boldsymbol{val}(t)$
%% whose vertices are those specified by the nullary symbols of $t$. Its \emph{%
%% width} is the number of labels that occur in $t$.

Using a standard convention, we will denote in the same way a function
symbol and the graph operation it defines. Hence, $relab_{a\rightarrow b}(t) 
$ \emph{is }a term if $t$ is a term and $relab_{a\rightarrow b}(G)$ \emph{%
denotes} a vertex labelled graph if $G$ denotes a vertex labelled graph.

We denote by $t/u$ the \emph{subterm of} $t$ \emph{issued from position} $u$%
. In the above example of term $t$, we have $t/u=\boldsymbol{1}(w)\oplus 
\boldsymbol{3}(w^{\prime })$ if $u$ is the first position that is an
occurrence of $\oplus $. Since no nullary symbol has two occurrences in a
term, for any two positions $u$ and $u^{\prime }$ in any term $t$, we have $%
t/u\neq t/u^{\prime}$.

The \emph{clique-width} of a graph $G$ without labels, denoted by $cwd(G),$
is the least width of a term $t$ that denotes some vertex labelling of $G$.
Such a term is said to be \emph{optimal}.

We will consider algorithms about graphs that are given by defining terms.
The time computations of these algorithms will depend on the widths of the
input terms. It is thus better to specify the input graphs by optimal terms
(see however the conclusion). But, deciding whether $cwd(G)\leq k$ is true
is an NP-complete problem (if $k$ is part of the input \cite%
{Fellows2009CliqueWidthIN}).

\section{Main result}

\begin{theorem}
Given a clique-width term of width $k$ and $m$ operation symbols (including nullary ones),
 $c$-colourability of $G$ can be checked in time 
$O(m\times{}min((c+1)^{2^k}\times{}2^{2k}\times{\log{}c},(2^{c}-2)^k\times{}c^3)\times{}k^3)$.
A more precise formula is given after analyzing the algorithm.
\end{theorem}

\begin{remark}
For $c=3$ we obtain $O(m\times{}6^k\times{}k^3)$,  matching the $%
        O^{*}((2^c-2)^k)$ bound \cite{DBLP:conf/icalp/Lampis18}.
        Here $O^{*}()$ denotes $O()$ up to factors polynomial
        in $k$, $c$, and $n$.
        For $c=\Theta(n)$ we obtain 
$O(m\times{}(n+1)^{2^k}\times{}2^{2k}\times{}k^3)$,  matching
the $n^{O(2^k)}$ bound \cite{DBLP:conf/wg/MakowskyRAG06}. Moreover, the
algorithm provided in  \cite{DBLP:conf/wg/MakowskyRAG06} contains an
operation with worst-case complexity $\Omega^{*}(n^{3\cdot{}2^{k}})$. We obtain
better worst case complexity by generalising an optimisation from  \cite%
{DBLP:conf/icalp/Lampis18}.
\end{remark}

\begin{remark}
If a graph has $n$ vertices,
a representing term has $n$ nullary operations for vertex addition
and $n-1$ union operations.
Furthermore, between two union operations or after the last union operation,
it is easy to optimise the unary operations
to first add edges in $O(k^2)$ operations,
then relabel in $O(k)$ operations,
as after any relabelling there is an unused label
that can be used as a buffer.
Thus $m=O(n\cdot k^2)$, and also $m-n=\Omega(n)$.
\end{remark}

Our proof is based on an algorithm using the optimisations from the cited
constructions. On the base level, the algorithm enumerates the possible
configurations of colours corresponding to labels, computing this for all
the subterms from the nullary symbols up to the entire term. The graph is
colourable if we find any valid configuration for the entire term.

There are two main optimisations for the low colour count case \cite%
{DBLP:conf/icalp/Lampis18}. The first optimisation is faster enumeration of
colourings of a union of two graphs by considering all overapproximations of
the sets of used colours for each label. This allows to look for 
identical label-colour relations for 
the two subgraphs instead of choosing compatible pairs of relations. The
complexity of the operation becomes linear in the number of possible
label-colour relations, instead of quadratic or cubic with other approaches.
Unlike \cite{DBLP:conf/icalp/Lampis18}, we implement this approach in a way
compatible with large numbers of colours.
The second optimisation is based on lookahead, namely computing which
edge addition operations outside a subterm affect the vertices corresponding
to the nullary symbols inside the term.
We call such additional small pieces of data computed for each subterm 
before starting the main part of the algorithm
annotations, following \cite{DBLP:journals/japll/CourcelleD12}.
The algorithm uses the fact that if
the vertices with label~$a$ will get connected to some other vertices, there
must be at least one colour that is not used for any vertex with the 
label~$a$.

The main optimisation for the case of a large number of colours is
identification of colourings differing only in permutation of colours.
It is quite similar to the optimisation used in  \cite{Kobler2003EdgeDS,%
DBLP:conf/wg/MakowskyRAG06}.
This
is achieved by storing for each set of labels the number of colours
corresponding to these labels.
We hope that our presentation is closer to the basic approach and 
thus more natural than the previous presentation of the case 
of large number of colours.

We now proceed to define the algorithm precisely.

\section{Algorithm}

The algorithm takes as the input a clique-width term $t$ of width $k$ 
with $n$ nullary symbols, and the number of colours $c$. 
In this section we assume that the term $t$
and the number of colours $c$ are fixed.
We will only consider subterms $t/u$ of $t$.

We start with an outline of the algorithm.
The basic approach is to compute the set of all possible colourings 
for each subterm, starting with the nullary operations in the leaves 
and going towards the root; 
for each operation we compute the set of colourings 
based on the colourings of the arguments.
As all the vertices sharing the same label
also share all the edges left to account for,
the colouring is stored as sets of labels for each colour.
To avoid unnecesasary repetition of the work,
we only store the colourings up to permutation of the colours,
i.e. as a multiset of set of colours.
To reduce the effort spent on the dead-ends,
we apply the constraints related to the edges
as soon as both vertices appear within the subterm
after a $\oplus$ operation
without waiting for the $add$ operation.
Moreover, once all the edges related to some label have been processed,
we stop keeping track of this label.
To optimise the union operation we observe that the colouring for the entire graph
adds some colours to some labels in comparison to the colourings of each subgraph.
We call this overapproximation and observe that finding a common overapproximation
can be done by computing all overapproximations of the schemes on both sides
and then computing the intersection.

We now proceed with a more detailed definition.
The main notion in the algorithm is that of a \emph{label-colouring scheme},
or simply a \emph{scheme}.
Such a scheme is a multiset of sets of labels.
Rougly speaking, a scheme denotes the number of
colours corresponding to some sets of labels.
For example, if we care about labels $a$ and $b$,
let us assume that the vertices with label $a$
are of colours $1,2,3$
and vertices with label $b$ are of colours $1$ and $2$.
This means that colours $1$ and $2$ are used for labels $a$ and $b$,
while the colour $3$ is only used for $a$.
The scheme for that case would be $\{2:\{a,b\},\{a\}\}$.

The algorithm aims to
compute the set of all the schemes describing the colourings of $\boldsymbol{%
val}(t)$ and check if there are any.

Before performing the main part of the work, the algorithm computes some
data we call annotations for each subterm.
 These annotations are described using the
following notions.

\begin{definition}
A label $a$ is \emph{used} in a subterm $t/u$ if some vertex of the labeled
graph $\boldsymbol{val}(t/u)$ has label $a$.

A label $a$ is \emph{a boundary label} in a subterm $t/u$ 
of a term $t$
 if there is a
vertex $x$ labeled $a$ in $\boldsymbol{val}(t/u)$  and there is a vertex $y$
in $\boldsymbol{val}(t)$  connected with $x$ in $\boldsymbol{val}(t)$,  such
that $y$ is not in $\boldsymbol{val}(t/u)$.  In other words, a label is a
boundary label if  there are some edges from a vertex with this label  to
vertices added later.

%% A label $a$ is \emph{live} in a subterm $t/u$
%% if there is a vertex $x$ labeled $a$ in $\boldsymbol{val}(t/u)$
%% and there is a vertex $y$ in $\boldsymbol{val}(t)$
%% connected with $x$ in  $\boldsymbol{val}(t)$,
%% such that $y$ is either outside  $\boldsymbol{val}(t/u)$
%% or not connected with $x$ in  $\boldsymbol{val}(t/u)$.
%% In other words, a label is live if there are some edges
%% to add later to some vertex with this label.

A pair of labels $(a,b)$ is \emph{pending }in a subterm $t/u$ if there are
non-connected vertices $x$ and $y$ in $\boldsymbol{val}(t/u)$ with labels $a$
and $b$ and there is an edge between $x$ and $y$ in $\boldsymbol{val}(t)$.
In other words, a pair of labels is pending if there are edges to be added
later (above in the syntactic tree) between some vertices with these labels.
\end{definition}

\begin{example}
        Consider the term 
        $relab_{1\rightarrow{}3}(add_{1,3}(add_{1,2}(\boldsymbol{1}(x)
        \oplus{}\boldsymbol{2}(y))
        \oplus{}\boldsymbol{3}(z)))$.
        Each nullary symbol as a subterm has its label as the only used (and boundary) label,
        and no pending pairs.
        The subterm $\boldsymbol{1}(x)\oplus{}\boldsymbol{2}(y)$
        has a pending pair $(1,2)$, and only $1$ is a boundary label
        (as $y$ is only connected to $x$ also defined inside the same subterm).
        Of course both $1$ and $2$ are used labels.
        The subterm $add_{1,2}(\boldsymbol{1}(x)\oplus{}\boldsymbol{2}(y))
        \oplus{}\boldsymbol{3}(z)$ has two pending pairs, $(1,2)$ and $(1,3)$
        and no boundary labels 
        as it contains all the vertices of the containing term.
        The entire term naturally has no boundary labels, and after the relabeling
        only $2$ and $3$ are used labels.
\end{example}

For each position $u$ in the term $t$, the algorithm precomputes the sets of
used labels, of boundary labels, and of pending pairs of labels.

We now define formally the notion of scheme.

\begin{definition}
Assume that we have a proper colouring of $G=\boldsymbol{val}(t)$ and
consider $\boldsymbol{val}(t/u)$.

Let $B(u)$ be the set of boundary labels of $t/u$.

For each colour $i$, let $C(i)$ the set 
of boundary labels $a\in{}B(u)$ such that
some vertex has label $a$ and colour $i$.
(Cf $R_{u}$ in introduction).

A \emph{scheme} is a multiset of sets of labels.
The \emph{description} of the chosen colouring at $u$
is the multiset of all $C(i)$.

For a scheme $s$ and a set of labels $L$ we let $s(L)$
denote the multiplicity of $L$ in $s$,
in particular, when $L$ is not in $s$ we have  $s(L)=0$.
\end{definition}

\begin{remark}
No boundary label $a$ can occur in all sets $C(i)$
for some $u$
(because no
colour would be left for the vertices that become 
adjacent to the $a$-vertices later on).

The sum of all the multiplicities in the description
of a colouring is always $c$, as each colour
adds one set (or increases its multiplicity).
A scheme describing a colouring also describes many
others, at least those obtained by permuting colours
but possibly more.
\end{remark}

\begin{example}
We let
$\{2:\{a,b\},\{b,c\},3:\varnothing \}$ denote the following scheme:
twice $\{a,b\}$, once $\{b,c\}$ and three colours do not colour any vertex
with a boundary label. As expected, we have $c=2+1+3=6$.
\end{example}

The main part of the algorithm computes a set of schemes 
for every position $u$ in $t$ based on the subterm $t/u$.
The schemes contain as elements some
sets of boundary labels. The aim is that every valid colouring of $%
\boldsymbol{val}(t)$ restricted to $\boldsymbol{val}(t/u)$ is described by
some scheme for $t/u$. Of course we do not guarantee that each scheme
describes a restriction of some valid colouring of $\boldsymbol{val}(t)$.
For the sake of simplicity, we only require that each scheme
overapproximates a scheme describing some valid colouring of $\boldsymbol{val%
}(t/u)$ in the following sense.

\begin{definition}
A \emph{single-step overapproximation} of a scheme $s$ is obtained by
reducing the multiplicity of some set of labels $L$ in $s$ by one, by
picking some label $a$, and by increasing the multiplicity of the set $L\cup
\{a\}$ by one. 
Additionally, the resulting scheme must still have a set of
labels with non-zero multiplicity not containing $a$. An \emph{%
overapproximation} of a scheme $s$ is any scheme reachable from $s$ by some
number of single-step overapproximations.
\end{definition}

\begin{remark}
For a description of some label-colour relation, a single-step overapproximation
describes the result of adding one more colour corresponding to the label $a$.

Overapproximation increases the \emph{weight} of a scheme
$\{...,m_{i}:L_{i},...\}$ 
        defined as the sum of the 
        $m_{i}.(1+\left\vert{}L_{i}\right\vert )$.
\end{remark}

The algorithm ensures that each scheme
in the set for $t/u$ overapproximates some
scheme describing a valid colouring of $\boldsymbol{val}(t/u)$. The
computation is defined recursively based on the symbol at position $u$.

For a nullary symbol with a label $a$, the algorithm returns $%
\{\{\{a\},(c-1):\varnothing\}\}$. In other words, there is just one
scheme, and it contains one colour corresponding to the set of labels $\{a\}$
(and $c-1$ unused colours).

For the edge addition between labels $a$ and $b$, the same set is used as
computed for the subterm that is the only argument of the operation.

For label renaming from $a$ to $b$ we modify the set of schemes computed for
the only argument of the operation. First for each scheme the algorithm
modifies each element by replacing $a$ with $b$ in the set (if present). If $%
b$ is already present, $a$ is just removed; if $a$ is not present, no change
is applied. In the process some of the label sets can become equal; in this
case their multiplicities are added together. For example, $\{\{a\},\{b\}\}$
becomes $\{2:\{b\}\}$. Afterwards, the algorithm removes all the
schemes with every element containing $b$. In other words, if $b$ is a
boundary label it cannot use all the colours.

The most elaborate operation is union.

In this case the subterm $t/u$ has two subterm arguments, $t/u_{1}$ and $%
t/u_{2}$. Let the sets of schemes corresponding to these subterms be $S_{1}$
and $S_{2}$. These sets of schemes describe some of the valid colourings of $%
\boldsymbol{val}(t/u_{1})$ and $\boldsymbol{val}(t/u_{2})$, including all
restrictions of valid colourings of $\boldsymbol{val}(t)$. The aim is to
enumerate the schemes corresponding to the valid colourings of $\boldsymbol{%
val}(t/u)$ (with all the restrictions of valid colourings of $\boldsymbol{val%
}(t)$ included). This is performed via the following steps. First, the
overapproximations are computed. The algorithm computes all
overapproximations of schemes in $S_{1}$ and $S_{2}$ that use all the
boundary labels of $u/t_{1}$ and $u/t_{2}$ and no other labels. This is done
by depth-first search using single-step overapproximation as edge relation.
More precisely,
%% if some labels are missing, we take only steps adding the
%% smallest missing label, and otherwise we make arbitrary single-step
%% overapproximations. 
if some labels are missing
we first add the missing labels one by one in a fixed
order. There are multiple ways to add each given label,
and we consider all of them.
Once all labels are used in the scheme, we consider
all the possible single-step overapproximations.
This yields sets $\overline{S_{1}}$ and $\overline{S_{2}}$. 
Remember that by definition of overapproximation,
as long as the initial schemes do not contain any labels
using all the colours, the same is true for the overapproximations.
Then $\overline{S_{1}}\cap \overline{S_{2}}$ is computed. In other words,
all the schemes simultaneously overapproximating some schemes from $S_{1}$
and $S_{2}$ are enumerated. Next schemes with colouring violation, i.e. some
element set containing both labels of some pending pair for $t/u$, are
removed.
The pending pairs are obtained from the annotations.
Then for each remaining schemes all labels that are not boundary
labels for $t/u$ are removed from all the elements of the scheme.
Boundary labels have also been precomputed as a part of the annotations.
If two elements of the scheme become the same set, their
multiplicities are combined. The resulting set of schemes is returned.

Consider an example where we have the sets of schemes
$\{\{2:\{a\},\varnothing{}\}\}$
and
$\{\{\{b\},2:\varnothing{}\}\}$, a pending pair $(a,b)$,
and only $b$ is a boundary label after the union
(i.e. after the union vertices with label $b$ have some 
external connections but vertices with the label $a$ 
do not).
Among the overapproximations of the scheme from the first set
there are 
$\{2:\{a\},\varnothing\}$,
$\{\{a\},\{a,b\},\varnothing\}$,
$\{2:\{a\},\{b\}\}$,
as well as some others.
But, for example, 
$\{2:\{a\},\{a,b\}\}$
is not an overapproximation as $a$ is in every set.
Of the above mentioned overapproximations,
$\{\{a\},\{a,b\},\varnothing\}$ 
and
$\{2:\{a\},\{b\}\}$
are common overapproximations for both sets of schemes.
The former, however, is removed for colouring violation.
Removing $a$ in the elements of the latter 
($a$ is not a boundary label)
yields the scheme
$\{2:\varnothing,\{b\}\}$.
Actually, this is the only scheme we would obtain
if we checked all the overapproximations.
The resulting set of schemes is
$\{\{2:\varnothing,\{b\}\}\}$.

Once the sets of schemes corresponding to all subterms of $t$ are
calculated, the algorithm verifies if the set of schemes for the term $t$ is
empty. A valid colouring of $\boldsymbol{val}(t)$ exists if and only if the
set of schemes for the term $t$ is not empty.

\section{Correctness}

Consider a subterm $t/u$.
Consider some colouring of the graph $\boldsymbol{val}(t/u)$.
A \emph{boundary set} of a colour $Q$
is the set of all boundary labels of $t/u$
having at least one vertex of the colour $Q$.
A \emph{description} of the colouring  is a scheme 
such that each set of boundary labels $L$
has multiplicity equal to the number of colours 
with boundary set equal to $L$.

We prove by induction on subterm structure that all schemes for each subterm 
$t/u$ are overapproximations of descriptions of some valid colourings of the
subgraph $\boldsymbol{val}(t/u)$, and descriptions of restrictions of all
valid colourings of the entire graph $\boldsymbol{val}(t)$ belong to the set
of schemes computed for $t/u$. More precisely, each scheme for subterm $t/u$
is an overapproximation of some scheme describing a valid colouring of the
induced subgraph in the full graph $\boldsymbol{val}(t)$ for the vertices of 
$\boldsymbol{val}(t/u)$.

In case of a single vertex there is just a single label and a single colour
used. It is clear that the condition holds.

In case of an edge addition, the set of vertices does not change and the
edge has been already taken into account before by induction assumption.

In case of relabeling from $a$ to $b$ there are two cases. Consider that the
same future edge additions apply to the vertices labeled $a$ and $b$, as
they share a label after relabeling. Therefore either $b$ is a boundary
label after relabeling, and then both $a$ and $b$ are boundary labels before
the relabeling; or $b$ is not a boundary label and then neither is $a$. If
neither $a$ nor $b$ are boundary labels, the relabeling does not change the
descriptions of any colourings, and the algorithm does not change the set of
schemes. If $a$ and $b$ are boundary labels, it is straightforward to verify
that the for each colouring of $\boldsymbol{val}(t/u)$ taking a describing
scheme before relabeling and applying the label replacement to the scheme
provides the same result as relabeling the graph first then taking the
description. It remains to show that the schemes we drop are not
descriptions of colourings of the entire graph $\boldsymbol{val}(t)$. But
indeed a boundary label cannot have vertices of all colours. Note that we
might remove overapproximations of descriptions of some valid colourings,
but we do not promise to keep all the overapproximations.

The last case is the union operation. Consider a colouring of $\boldsymbol{%
val}(t/u)$. Each colour is used for some labels. The same colour is used for
two subsets of labels in $\boldsymbol{val}(t/u_1)$ and $\boldsymbol{val}%
(t/u_2)$, Thus we can find a common overapproximation for the descriptions
of the two restrictions. The condition that an overapproximation cannot make
a label correspond to all the colours is not violated, as each label we care
about is either a boundary label for $t/u$ and does not use all the colours,
or a boundary label in $t/u_1$ or $t/u_2$ but not in $t/u$ and is in some
pending pair of labels, thus not using any colours used by the second label
in the pair. Moreover, the description of this colouring only differs from
this common overapproximation by dropping non-boundary labels in each
element of the scheme.

Now we show that each scheme $s$ that is not removed is an overapproximation
of the description of some colouring of $\boldsymbol{val}(t/u)$. Consider
the scheme $\tilde{s}$ such that $s$ was obtained by ignoring non-boundary
labels in $\tilde{s}$. The scheme $\tilde{s}$ was obtained from two such
schemes $s_1$ and $s_2$ for the subgraphs $\boldsymbol{val}(t/u_1)$ and $%
\boldsymbol{val}(t/u_2)$. Without loss of generality assume that $s_1$ and $%
s_2$ are exact descriptions are not overapproximations, as overapproximation
is transitive. Consider some valid colourings $C_1$ and $C_2$ of $%
\boldsymbol{val}(t/u_1)$ and $\boldsymbol{val}(t/u_2)$ with descriptions $s_1
$ and $s_2$. Such colourings exist by induction assumption. Picking an
arbitrary colour to be assigned to the additional label for each single-step
overapproximation, we can give each colour a set of labels larger than the
set of corresponding labels according to $C_1$ so that each set of labels $L$
is used $s(L)$ times. The same applies to $C_2$. As the colours can be
permuted, we assume without loss of generality that we obtain the same
function from colours to sets of labels. In this case we can just unite the
colourings $C_1$ and $C_2$ and obtain a valid colouring of induced subgraph
corresponding to the vertices of $\boldsymbol{val}(t/u)$. Indeed, each edge
in this subgraph either connects the vertices on the same side of the union
(then the colours must be different as $C_1$ and $C_2$ are valid), or on
different sides. In the latter case the corresponding labels are boundary
labels in $t/u_1$ and $t/u_2$, and this pair of labels is pending for $t/u$,
thus $\tilde{s}$ would have been removed. Note that this case analysis is
per edge, not per pair of labels; the same pair of labels might correspond
to edges of both types.

It is straightforward to verify that the description of the restriction of a
colouring of the full graph $\boldsymbol{val}(t)$ will not be removed, as it
cannot have boundary labels using all the colours nor pending pairs sharing
a colour.

We conclude that the algorithm indeed computes some set of overaproximations
of descriptions of colourings. It remains to observe that for the full term $%
t$ we have some overapproximations of descriptions of colourings, including
all exact descriptions of colourings. This set is empty iff there is no
colouring. This concludes the proof of correctness.

\begin{remark}
The correctness proof is essentially constructive.
In other words,
if we have a scheme for a term and the corresponding schemes for all the subterms,
the correctness proof explains how to combine
the trivial colourings of
single-vertex graphs represented by the leaf subterms
into a colouring for the entire graph.

To be able to obtain intermediate schemes,
we can
store the first justification of inclusion 
for each intermediate scheme computed during the algorithm.
\end{remark}

\section{Complexity}

We use sequences of boolean values of length $k$ to represent sets of
labels. We use PATRICIA trie data structure \cite{10.1145/321479.321481} on
strings of length $k$ with integer labels in leaves to represent schemes,
and also PATRICIA tries for some string encoding of schemes to represent
sets of schemes (any set data structure on strings with operations taking
time proportional to string traversal time is suitable here). We store the
pointers to the arguments and to the parent operation for each position in
the term (this is trivial to compute in linear time).

The annotations can be precomputed in the time $m \times{} k^2$.

Each scheme has at most $c$ entries, as we do not need to keep elements with
zero multiplicity, and at most $2^k$ entries as this is the number of sets
of labels. The total number of schemes using $k^{\prime}$ labels is bounded
both by ${\binom{{c+2^k-1}}{{2^k}}}\leq{}(c+1)^{2^k}$ counting them as
ordered partitions of $c$ into $2^k$ summands, and by $(2^c-2)^k$ as each
label has a non-empty set of colours and also cannot use all the colours.
Let $D_s$ be $min(c,2^k)$ and $N_s$ be $min((c+1)^{2^k}, (2^c-2)^k)$. When
encoding a scheme, each set of labels needs $l_s = k + \log c$ bits.

For the main computation, it is easy to see that the union operation is the
most expensive one. The overapproximations are computed by traversing a
graph. The graph can be stratified by the set of labels used; for each size
of the label set there is at most one possible set of used labels (as we add
labels in a fixed order). It is clear that the number of schemes for the
largest set of labels is larger than the total number for the smaller sets,
so we bound the number of vertices by $2N_s$. The degree of the nodes in the
graph is at most $D_s\times{}k$, as we pick a set of labels with at least
one colour, then pick a label to add. Overall the traversal takes $2 N_s D_s
k$ operations on schemes. Intersection takes just $3 N_s$ operations on
schemes considering lookup/insertion on sets of schemes an operation on
schemes. Removal of non-boundary labels takes at most $N_s k$ operations on
schemes. Removing invalid schemes takes $N_s k^2$ operations on schemes, as
we need to consider up to $\frac{k(k-1)}{2} + k$ reasons to remove a scheme;
the reason can be either an overflow in a label or a pending pair of labels.
Each operation on schemes takes time at most proportional to $D_s l_s$
steps. The total time is $O(N_s (D_s+k) k \times D_s l_s)$.

Given that we have $m$ operations, precomputations are much cheaper than the
main part, and the total runtime of the algorithm is $O(m\times{}%
min((c+1)^{2^k}, (2^c-2)^k)\times{}(min(c,2^k)+k)\times{}k\times{}min(c,2^k)%
\times{}(k+\log c))$.

\section{Conclusion and further work}

We have shown that a single algorithm can verify $c$-colourability of a
graph provided as a clique-width decomposition with performance matching the
state of the art both in the few-colours and many-colours cases.

A natural question arises whether a modification of this algorithm can be
used to count the number of possible colourings. We conjecture that this can
be done without significantly exceeding the complexity of the cited
algorithms from the literature.

As the most complicated and the most expensive operations 
is the union operation,
it might be of interest to see what simplifications and
optimisations can be achieved for linear clique-width terms,
i.e. when the second subgraph in each union has exactly
one vertex.

The algorithm presented optimises the worst case complexity of computing the
sets of schemes. Empirical evidence shows that for some graphs there are few
schemes corresponding to each subterm, making the approach based on
precomputing all overapproximations less efficient than a naive approach
based on enumerating all pairs of schemes. In this approach, for a pair of
schemes we solve a matching-like problem to figure out which colour used for
colouring the first subgraph correspond to which colour in the colouring of
the second subgraph. One can use either a brute force approach or some
algorithms based on maximum flow and similar considerations (e.g.  \cite%
{DBLP:conf/isaac/Uno97}) to enumerate the possible matchings. We believe it
is an interesting question to find some natural class of graphs where such
an approach guarantees a better upper bound on the runtime.

\section{Acknowledgements}

We are grateful to Michael Lampis for interesting discussions. 
 We are grateful to the anonymous reviewers
 for their feedback on the presentation.

\bibliographystyle{plain}
\bibliography{references}

\end{document}